\documentclass[twocolumn,trackchanges]{aastex631}

\usepackage{natbib}
\usepackage{graphicx}
\usepackage{epstopdf}
\usepackage{amsmath}
\usepackage{hyperref}
\hypersetup{hidelinks,
	colorlinks=true,
	allcolors=black,
	pdfstartview=Fit,
	breaklinks=true}
\let\oldAA\AA
\renewcommand{\AA}{\text{\normalfont\oldAA}}

\begin{document}

\title{XMM-Newton Observations of Two Archival X-ray Weak Type 1 Quasars: Obscuration Induced X-ray Weakness and Variability}

%\correspondingauthor{Bin Luo}
%\email{bluo@nju.edu.cn}

\author[0000-0002-2420-5022]{Zijian Zhang}
\affiliation{School of Astronomy and Space Science, Nanjing University, Nanjing 210093, China}
\affiliation{Key Laboratory of Modern Astronomy and Astrophysics (Nanjing University), Ministry of Education, China}

\author[0000-0002-9036-0063]{Bin Luo}
\affiliation{School of Astronomy and Space Science, Nanjing University, Nanjing 210093, China}
\affiliation{Key Laboratory of Modern Astronomy and Astrophysics (Nanjing University), Ministry of Education, China}

\author[0000-0002-0167-2453]{W. N. Brandt}
\affiliation{Department of Astronomy \& Astrophysics, 525 Davey Lab, The Pennsylvania State University, University Park, PA 16802, USA}
\affiliation{Institute for Gravitation and the Cosmos, The Pennsylvania State University, University Park, PA 16802, USA}
\affiliation{Department of Physics, 104 Davey Lab, The Pennsylvania State University, University Park, PA 16802, USA}

\author[0000-0002-5830-3544]{Pu Du}
\affiliation{Key Laboratory for Particle Astrophysics, Institute of High Energy Physics, Chinese Academy of Sciences, 19B Yuquan Road, Beijing 100049, China}

\author{Chen Hu}
\affiliation{Key Laboratory for Particle Astrophysics, Institute of High Energy Physics, Chinese Academy of Sciences, 19B Yuquan Road, Beijing 100049, China}

\author[0000-0002-9335-9455]{Jian Huang}
\affiliation{School of Astronomy and Space Science, Nanjing University, Nanjing 210093, China}
\affiliation{Key Laboratory of Modern Astronomy and Astrophysics (Nanjing University), Ministry of Education, China}

\author[0000-0003-3349-4855]{Xingting Pu}
\affiliation{College of Science, Nanjing Forestry University, Nanjing, Jiangsu 210037, People’s Republic of China}

\author[0000-0001-9449-9268]{Jian-Min Wang}
\affiliation{Key Laboratory for Particle Astrophysics, Institute of High Energy Physics, Chinese Academy of Sciences, 19B Yuquan Road, Beijing 100049, China}

\author[0000-0001-9314-0552]{Weimin Yi}
\affiliation{Yunnan Observatories, Chinese Academy of Sciences, Kunming, 650216, China}
\affiliation{Department of Astronomy \& Astrophysics, 525 Davey Lab, The Pennsylvania State University, University Park, PA 16802, USA}



\begin{abstract}
We report \hbox{XMM-Newton} observations of two examples of an unclassified type of \hbox{X-ray} weak quasars from the \citet{2020ApJ...900..141P} survey of \hbox{X-ray} weak quasars in the Chandra archive, SDSS J083116.62+321329.6 at $z=1.797$ and SDSS J142339.87+042041.1 at $z=1.702$. They do not belong to the known populations of \hbox{X-ray} weak quasars that show broad absorption lines, weak ultraviolet (UV) broad emission lines, or red optical/UV continua. Instead, they display typical quasar UV spectra and spectral energy distributions. In the \hbox{XMM-Newton} observations, both quasars show nominal levels of \hbox{X-ray} emission with typical quasar \hbox{X-ray} spectral shapes (\hbox{power-law} photon indices of $1.99^{+0.27}_{-0.23}$ and $1.86^{+0.15}_{-0.14}$), displaying strong \hbox{X-ray} variability compared to the archival Chandra data (variability factors of $4.0^{+1.6}_{-1.4}$ and $9.0^{+7.4}_{-3.8}$ in terms of the 2 keV flux density). Simultaneous optical (rest-frame UV) spectra indicate no strong variability compared to the archival spectra. Long-term optical/UV and infrared light curves do not show any substantial variability either. We consider that the \hbox{X-ray} weakness observed in the Chandra data is due to \hbox{X-ray} obscuration from a small-scale dust-free absorber, likely related to accretion-disk winds. Such \hbox{X-ray} weak/absorbed states are probably rare in typical quasars, and thus both targets recovered to \hbox{X-ray} nominal-strength states in the \hbox{XMM-Newton} observations.

\end{abstract}

\keywords{galaxies: active – quasars: individual – X-rays: galaxies}
\section{Introduction}
\label{sec:intro}
Quasars are powered by accretion onto supermassive black holes (SMBHs) in the centers of massive galaxies. Luminous \hbox{X-ray} emission is a ubiquitous property of quasars, which is believed to originate largely from the “corona” located around the inner accretion disk via Comptonization of optical/ultraviolet (UV) seed photons \citep[e.g.,][]{2010arXiv1008.2287D,2014SSRv..183..121G,2017AN....338..269F,2023arXiv230210930G}. A significant correlation has been observed between the coronal \hbox{X-ray} emission and the \hbox{accretion-disk} optical/UV emission, typically expressed as the relation between the X-ray-to-optical \hbox{power-law} slope parameter ($\alpha_{\rm OX}$)\footnote{$\alpha_{\rm OX}$ is defined as $\alpha_{\rm OX}=-0.3838 \log(f_{\rm 2500 \AA}/f_{\rm 2 keV})$, where $f_{\rm 2500 \AA}$ and $f_{\rm 2 keV}$ are the rest-frame 2500 $\rm \AA$ and 2 keV flux densities, respectively.} and the 2500 $\rm \AA$ monochromatic luminosity ($L_{\rm 2500 \AA}$), over $\approx$ 5 orders of magnitude in UV luminosity \citep[e.g.,][]{2006AJ....131.2826S,2007ApJ...665.1004J,2017A&A...602A..79L,2020ApJ...900..141P}. A quasar is considered to be \hbox{X-ray} weak if it deviates below the $\alpha_{\rm OX} \textrm{--} L_{\rm 2500 \AA}$ relation, showing weaker than expected \hbox{X-ray} emission. The amount of \hbox{X-ray} weakness is often quantified by the $\Delta \alpha_{\rm OX}$ parameter, defined as the difference between the observed and expected $\alpha_{\rm OX}$ values ($\Delta \alpha_{\rm OX}=\alpha_{\rm OX} - \alpha_{\rm OX,exp}$); the corresponding \hbox{X-ray} weakness factor is $f_{\rm weak} = 10^{-\Delta \alpha_{\rm OX}/0.384}$.

Except for a still-uncertain potential rare population of intrinsically \hbox{X-ray} weak quasars \citep[e.g.,][]{2007ApJS..173....1L,2007ApJ...663..103L,2014ApJ...794...70L,2018ApJ...859..113L,2022ApJ...936...95W}, observations of \hbox{X-ray} weak type 1 quasars are generally ascribed to \hbox{X-ray} obscuration. For example, broad absorption line (BAL) quasars are generally \hbox{X-ray} weak due to absorption by a clumpy outflowing wind or “shielding gas” associated with the wind \citep[e.g.,][]{1995ApJ...451..498M,2006ApJ...644..709G,2014MNRAS.438..604B,2016MNRAS.458..293M}. \citet{2020ApJ...900..141P} performed a systematic investigation of \hbox{X-ray} emission from a large sample of Sloan Digital Sky Survey (SDSS) non-BAL type 1 quasars using Chandra archival observations, and they found a population of non-BAL \hbox{X-ray} weak quasars. The fraction of quasars that are \hbox{X-ray} weak by factors of $\geq$ 6 is 5.8\% $\pm$ 0.7\%. They further classified these \hbox{X-ray} weak quasars into three categories based on their optical spectral features: weak emission-line quasars (WLQs), red quasars, and unclassified objects. Previous studies have revealed that the \hbox{X-ray} weakness of the former two types of quasars is likely due to \hbox{X-ray} obscuration \citep[e.g.,][]{2005ApJ...634..183W,2006AJ....132.1977H,2011ApJ...736...28W,2012ApJ...747...10W,2015ApJ...805..122L,2018MNRAS.480.5184N,2022MNRAS.511.5251N}.

The nature of the unclassified type of \hbox{X-ray} weak quasars in \citet{2020ApJ...900..141P} is uncertain (see discussion in Section 5.2.3 of \citealt{2020ApJ...900..141P}). They have UV continuum and emission-line spectra very similar to the SDSS quasar composite spectrum; i.e., there are no BALs or mini-BALs, weak emission lines, or redder than typical continua. They even have typical quasar spectral energy distributions (SEDs) from the infrared (IR) to UV. These quasars were serendipitously detected by Chandra, and they have at most a few tens of photons in the 0.5--7 keV band. The derived effective \hbox{power-law} photon indices are generally small ($\approx 1$) but with substantial uncertainties, suggestive of \hbox{X-ray} absorption. Moreover, the SDSS spectra and Chandra data were not obtained simultaneously, and they are separated by \hbox{$\approx$ 1--4 years} in the rest frame. It is thus probable that the observed \hbox{X-ray} weakness and typical UV spectra are due to variability effects. For example, there is a rare population of extremely \hbox{X-ray} variable quasars that have displayed strong \hbox{X-ray} variability (variability factors $\gtrsim$ 6) with no corresponding variability in the optical/UV spectra (e.g., PG 1211+143: \citealt{2009MNRAS.399..750B}; PG 0844+349: \citealt{2011MNRAS.412..161G,2012ApJ...746...54G}; SDSS J135058.12 +261855.2: \citealt{2022ApJ...930...53L}). They become significantly \hbox{X-ray} weak in their low states, and the \hbox{X-ray} weakness is often explained with absorption. Another example is the strong \hbox{long-term} \hbox{X-ray} variability that has been observed in several “\hbox{changing-look}” (showing type transitions) quasars \citep[e.g.,][]{2015ApJ...800..144L}. If the low-state \hbox{X-ray} fluxes and high-state optical/UV fluxes are mixed, the derived $\alpha_{\rm OX}$ values would appear smaller than those expected from the $\alpha_{\rm OX}\textrm{--}L_{\rm 2500\AA}$ relation. It is also possible that during the Chandra observations these quasars developed BALs which were not present in the earlier SDSS observations; a small population of quasars has been found to show emerging or disappearing BALs \cite[e.g.,][]{2012ApJ...757..114F,2017MNRAS.469.3163M,2018A&A...616A.114D,2018ApJ...862...22R,2019MNRAS.482.1121S,2021ApJS..255...12Y}. Nevertheless, additional \hbox{X-ray} and optical spectroscopic observations are needed to clarify the nature of these exceptional objects. 

%Besides, a fraction of BAL quasars can show emerging or disappearing BALs \cite[e.g.,][]{2012ApJ...757..114F,2017MNRAS.469.3163M,2018A&A...616A.114D,2018ApJ...862...22R,2019MNRAS.482.1121S,2021ApJS..255...12Y}, thus these unclassified type of X-ray weak quasars may have BALs during the Chandra epoch but not when observed by XMM-Newton.

In this study, we present additional deeper \hbox{XMM-Newton} observations of two examples of the unclassified type of \hbox{X-ray} weak quasars in \citet{2020ApJ...900..141P}, SDSS J083116.62+321329.6 and SDSS J142339.87+042041.1 (hereafter SDSS J0831+3213 and SDSS J1423+0420). At a redshift of 1.797, SDSS J0831+3213 was detected serendipitously by a Chandra observation on 2007 December 22. It has $17.4^{+5.9}_{-4.8}$  counts in the 0.5--7 keV band with an effective \hbox{power-law} photon index ($\Gamma_{\rm eff}$) of $1.0^{+0.6}_{-0.5}$ \citep{2020ApJ...900..141P}. SDSS J1423+0420 is at $z=1.702$. It was detected by a Chandra observation on 2012 December 15, with $6.0^{+3.9}_{-2.7}$ counts in the \hbox{0.5--7} keV band and $\Gamma_{\rm eff}<1.3$ \citep{2020ApJ...900..141P}. Using the $\alpha_{\rm OX}\textrm{--}L_{\rm 2500\AA}$ relation in \citet{2006AJ....131.2826S}, these quasars have $\Delta \alpha_{\rm OX}$ values of $-0.34^{+0.05}_{-0.05}$ and $-0.48^{+0.08}_{-0.10}$, corresponding to $f_{\rm weak}$ values of $7.7^{+2.9}_{-1.9}$ and $17.4^{+14.2}_{-6.8}$, respectively. In the deeper \hbox{XMM-Newton} observations, both quasars recovered to nominal levels of \hbox{X-ray} emission, and thus they displayed strong \hbox{X-ray} variability compared to the archival Chandra data. Optical spectra simultaneous to the XMM-Newton observations are also available. The results support the notion that these rare \hbox{X-ray} weak quasars were not peculiar objects having weak/suppressed coronal \hbox{X-ray} emission; instead, they were simply caught in unusual \hbox{X-ray} absorbed states.

%Motivated by the remaining uncertainties of the unclassified X-ray weak quasars in \citet{2020ApJ...900..141P}, we obtained additional deeper X-ray observations with simultaneous rest-frame UV spectroscopic observations of two of these quasars, i.e., SDSS J083116.62+321329.6 and SDSS J142339.87+042041.1 (hereafter SDSS J0831+3213 and SDSS J1423+0420), which allow us to understand the nature of their X-ray weakness better. In this paper, we report the transition to the X-ray nominal-strength state with no apparent optical/UV variability of these two quasars, which indicates that X-ray obscuration is the most probable reason for their X-ray weakness. 

The paper is organized as follows. We describe the X-ray and simultaneous optical spectroscopic observations in Section \ref{sec:xray-optical_observation}. The \hbox{X-ray} and multiwavelength properties of the two quasars are presented in Section \ref{sec:xray-multiwavelength}. We discuss and summarize our results in Section \ref{sec:summary_discussion}.
Throughout this paper, we use a cosmology with $H_{0}=67.4 \rm km~s^{-1}~Mpc^{-1}$, $\Omega_{\rm M} = 0.315$, and $\Omega_{\rm \Lambda} = 0.686$ \citep{2020A&A...641A...6P}. Measurement uncertainties are quoted at a 1$\sigma$ confidence level, while upper limits are quoted at a 90\% confidence level.

%The paper is organized as follows. We describe the X-ray and simultaneous optical spectroscopic observations and data analyses in Section \ref{sec:xray-optical_observation}. The X-ray and multi-wavelength properties of the two quasars are presented in Section \ref{sec:xray-multiwavelength}. %In Section \ref{sec:discussion}, we present that these two quasars do not belong to the two known X-ray weak non-BAL quasar catalogues in \citet{2020ApJ...900..141P}, and their X-ray weakness should be due to obscuration. The absorbers are likely clouds in the broad line region (BLR) or the clumpy dust-free wind launched from the accretion disk. We also discuss the amplitude and fraction of such quasars. 
%We summarize and discuss in Section \ref{sec:summary_discussion}. Throughout this paper, we use J2000 coordinates and a cosmology with $H_{0}~=~67.4~\rm km~s^{-1}~Mpc^{-1}$, $\Omega_{\rm M}$ = 0.315, and $\Omega_{\rm \lambda}$ = 0.686 \citep{2020A&A...641A...6P}. Uncertainties of all values are all at 1$\sigma$ confidence level. Galactic neutral hydrogen absorption level from \citet{HI4PI} was included in the X-ray spectral modelling. All the infra-red to UV photometric and spectroscopic data were corrected for extinction using the \cite{2019ApJ...886..108F} Milky Way extinction model with $R_{\rm V}$ = 3.1 and Galactic extinction $E({\rm B-V})$ at the source directions. The Galactic extinction values of SDSS J0831+3213 and SDSS J1423+0420 were 0.0409 and 0.0237, respectively, obtained from \cite{2011ApJ...737..103S}.

\section{X-ray and optical Observations} 
\label{sec:xray-optical_observation}

% \begin{table}
% \setlength\tabcolsep{5pt} 
% %\linespread{3} 
% % \renewcommand\arraystretch{1.5}
% \centering
% \caption{}
% \label{tab:target_properties}
% \begin{tabular}{cccc}
% \hline
% Name (SDSS J)&0831 + 3213&1423 + 0420\\
% \hline
% \hline

% z&1.787&1.702&\\
% $m_i$&18.7&18.8&\\
% SDSS Date&12/12/2002&05/20/2001&\\
% CXO Date&12/22/2007&12/15/2012&\\
% CXO Exp (ks)&10.0&10.0&\\
% Net Counts&$17.4^{+5.9}_{-4.8}$&$6.0^{+3.9}_{-2.7}$&\\ 
% $\Gamma_{\rm eff}$&$1.0^{+0.6}_{-0.5}$&$<1.3$&\\ 
% $\alpha_{\rm ox}$&$-2.0\pm0.1$&$-2.0\pm0.1$&\\
% $f_{\rm weak}$&$7\pm 1$&$14^{+3}_{-5}$&\\ 
% \hline
% \end{tabular}
% \end{table}
\begin{deluxetable*}{lcccccccc}
\tablecaption{Basic Object Properties and Their \hbox{XMM-Newton} Observations}
\tablehead{
\colhead{Object }  &
\colhead{$z$}  &
\colhead{$m_{B}$}  &
\colhead{$\log M_{\rm BH}$}  &
\colhead{$L/L_{\rm Edd}$} &
%\colhead{$\log L_{\rm 2500~{\textup{\AA}}}$} &
\colhead{$N_{\rm H, Gal}$}&
%\colhead{Observatory}	&
\colhead{Observation}	&
\colhead{Obs. Date} &
\colhead{Exposure Time} \\
\colhead{Name}	&
\colhead{}	&
\colhead{}	&
\colhead{($M_{\Sun}$)}	&
\colhead{}	&
%\colhead{(erg s$^{-1}$ Hz$^{-1}$)}	&
\colhead{($10^{20}$~cm$^{-2}$)}	&
%\colhead{} &
\colhead{ID}	&
\colhead{} &
\colhead{(ks)}\\
\colhead{(1)}         &
\colhead{(2)}         &
\colhead{(3)}         &
\colhead{(4)}         &
\colhead{(5)}         &
\colhead{(6)}         &
\colhead{(7)}         &
\colhead{(8)}		  &
\colhead{(9)}		  
%\colhead{(10)}		  &
%\colhead{(11)}		 
}
\startdata
%SDSS J0831+3213 &1.797	&18.7  &$9.32^{+0.08}_{-0.08}$	&$0.18^{+0.04}_{-0.03}$	&30.9  &3.87  	&\chandra	&9271	 &2007 Dec 22	&10.0	\\
SDSS J0831+3213 &1.797	&18.7  &$9.32$	&$0.18$ &3.87 &0861260101&2021 Apr 07&22.1	\\
%SDSS J1423+0420 &1.702	&18.8   &$9.37^{+0.06}_{-0.06}$	&$0.10^{+0.01}_{-0.01}$	&30.8 &2.07  	&\chandra 	&15375		&   2012 Dec 15 &10.0		\\
%		 &		&		&		&	&	&	&\chandra 	& 2985	& 2001 Dec 17     &9.8		\\
SDSS J1423+0420 &1.702	&18.8   &$9.37$	&$0.10$ &2.07  &0861260301&2021 Jan 25&36.4	
%		 &		&		&		&	&	&	&\hbox{XMM-Newton} 		&0761910201	& 2015 Nov 29	       &39.9	\\
%		 &		&		&		&	&	&	&\nustar 	&60101004002	& 2015 Nov 28	      &54.7
\enddata

\tablecomments{
Cols. (1) and (2): object name and redshift.
Col. (3): $B$--band magnitude.
Cols. (4) and (5): single-epoch virial SMBH mass and Eddington ratio estimates from \citet{2022ApJS..263...42W}.
%Col. (6): 2500~{\textup{\AA}} monochromatic luminosity from \citet{2020ApJ...900..141P}.
Col. (6): Galactic neutral hydrogen column density. 
Col. (7): \hbox{XMM-Newton} observation ID.
Col. (8): observation start date.
Col. (9): cleaned EPIC pn exposure time.
}
\label{tab:objectproperty}
\end{deluxetable*}

\subsection{XMM-Newton observations}
\label{subsec:target_property}
%SDSS J0831+3213 and SDSS J1423+0420 are two quasars from SDSS Seventh Data Release \citep[DR7;][]{2009ApJS..182..543A} with \textit{\chandra} coverage picked out as unclassified X-ray weak individuals by a systematic investigation of X-ray weak quasars populations \citep{2020ApJ...900..141P}, which have redshifts of 1.797 and 1.702, respectively. With effective \hbox{power-law} photon indices of $1.0^{+0.6}_{-0.5}$ and $\textless 1.3$, respectively, the two quasars are X-ray weak by factors of 8 and 18.1 compared to the $\alpha_{\rm OX} \textrm{--} L_{\rm 2500 \AA}$ relation according to \citet{2020ApJ...900..141P} (the 2500 $\rm \AA$ luminosities were converted from the absolute i-band magnitudes $M_{\rm i}(z~=~2)$). These two quasars were serendipitously detected by Chandra on 2007 December 22 and 2012 December 15, with exposure times of less than 10 ks. They have only $17.4^{+5.9}_{-4.8}$ and $6.0^{+3.9}_{-2.7}$ photons in the 0.5-7 keV band \citep{2020ApJ...900..141P}, respectively, which makes their spectral shapes very uncertain.

We obtained an \hbox{XMM-Newton} observation of SDSS J0831+3213 on 2021 April 07 with a total exposure time of 34.9 ks. SDSS J1423+0420 was observed on 2021 January 25 with a total exposure time of 57.5 ks. The \hbox{X-ray} data were processed using the \hbox{XMM-Newton}  Science Analysis System \citep[SAS v.20.0.0;][]{2004ASPC..314..759G} and the latest calibration files. All the EPIC pn, MOS1, and MOS2 data were used in our study. We reduced the pn and MOS data following the standard procedure described in the SAS Data Analysis Threads.\footnote{\url{https://www.cosmos.esa.int/web/XMM-Newton/sas-threads}.} Background flares were filtered to generate cleaned event files. The cleaned pn, MOS1, and MOS2 exposure times are 22.1, 32.6, and 32.3 ks for SDSS J0831+3213, and they are 36.4, 53.0, and 52.4 ks for SDSS J1423+0420. Both targets are significantly detected in the pn and MOS images. For each source, we extracted a source spectrum using a circular region with a radius of 30$\rm \arcsec$ centred on the optical source position. The total 0.3--10 keV spectral counts combining the pn and MOS spectra are 571 and 1163 for SDSS J0831+3213 and SDSS J1423+0420, respectively. For each source, a background spectrum was extracted from a few nearby circular source-free regions on the same CCD chip with a total area of about four times the area of the source region. Spectral response files were generated using the tasks {\sc rmfgen} and {\sc  arfgen}. We then group the source spectra with at least 25 counts per bin for spectral fitting.

The basic information for the two observations is listed in Table \ref{tab:objectproperty}. We also present in the table some basic object properties, including the Mg II-based single-epoch virial SMBH masses and Eddington ratios from \citet{2022ApJS..263...42W}. We note that these mass and Eddington ratio estimates have substantial uncertainties \citep[$\sim0.5$~dex; e.g.,][]{2012ApJ...753..125S}. The Galactic neutral hydrogen column densities from \citet{2016A&A...594A.116H} were used in the \hbox{X-ray} spectral analysis below.

%We note that the Mg II-based virial mass estimates may have a considerable scatter of >0.5 dex compared with the true mass \citep{2012ApJ...753..125S}. 

The \hbox{XMM-Newton} Optical Monitor (OM) observations used three UV filters, including UVM2, UVW1, and U, with effective wavelengths of 2310 $\rm \AA$, 2910 $\rm \AA$, and 3440 $\rm \AA$, respectively. The OM observation of SDSS J1423+0420 was heavily contaminated by scattered light from a nearby bright field star,\footnote{See an example in Section 2.10 of the OM calibration status document: \url{https://xmmweb.esac.esa.int/docs/documents/CAL-TN-0019.pdf}.} and we were not able to extract any useful photometry. For SDSS J0831+3213, the OM data were reduced using the task {\sc omchain}, and the photometric measurements of every exposure were recorded in the generated SWSRLI files. We extracted the magnitude measurements for each filter from these files and computed the mean magnitudes in the three OM bands. 

\begin{figure*}
\hspace{-0.2cm}
 \includegraphics[width=0.95\textwidth]{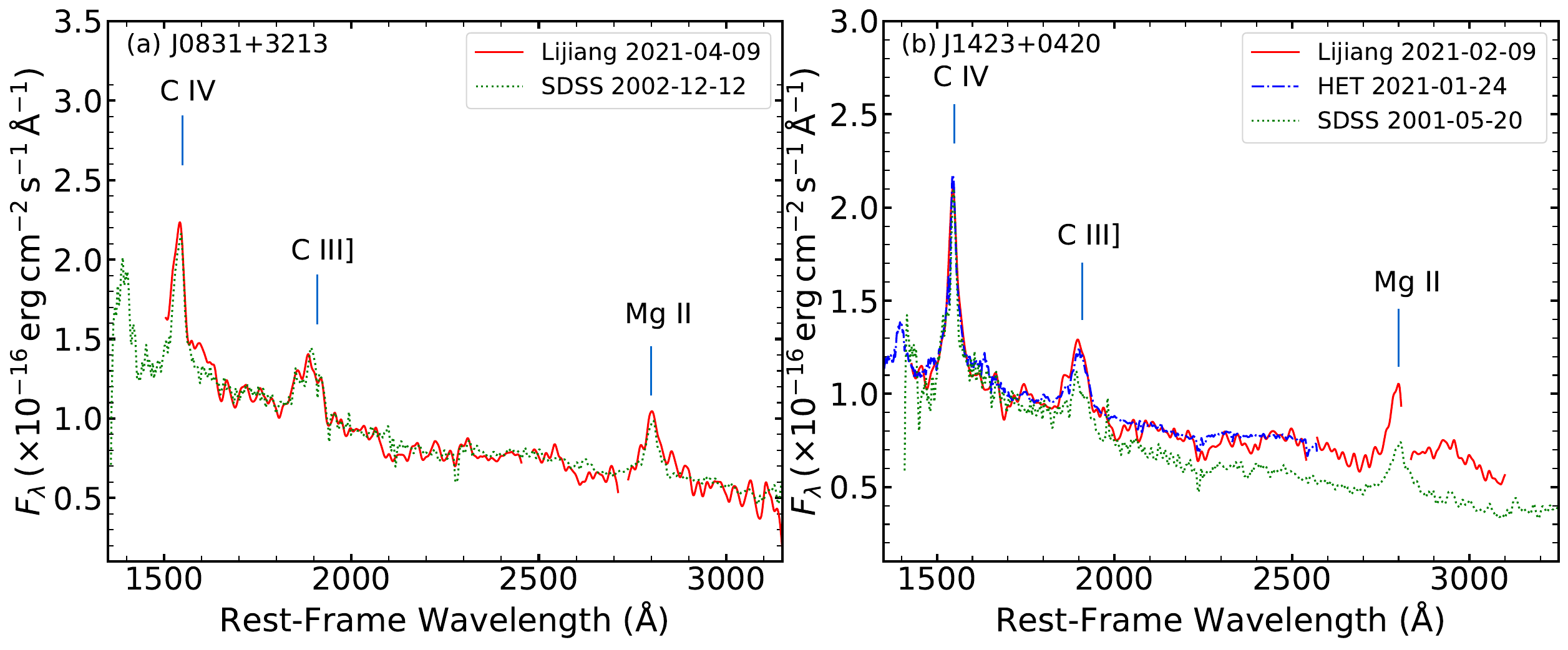}
 \centering
 \caption{(a) The Lijiang and SDSS spectra of SDSS J0831 + 3213. (b) The Lijiang, HET, and SDSS spectra of SDSS J1423 + 0420. The spectra were smoothed by a gaussian kernel with a standard deviation of 5 pixels. Some of the major UV emission lines are marked. \textcolor{black}{Two main telluric absorption windows (observed-frame 6864--6945 $\AA$ and 7586--7658 $\AA$) of the Lijiang spectra are masked out.} Overall, the rest-frame UV spectra of these two quasars did not vary greatly over the years. }
 \label{fig:optical_spectra}
\end{figure*}

\begin{deluxetable*}{ccccccc}
%\renewcommand\arraystretch{1.5}
%\begin{table}[t]{cccccccccccc}
%\tabletypesize{\scriptsize}
%\tabletypesize{\small}
\tablewidth{0.2pt}
\tablecaption{\hbox{XMM-Newton} Spectral Fitting Results
}
\tablehead{
\colhead{Object Name}&
%\colhead{Observation}&
%\colhead{Band} &
%\colhead{Total} &
%\colhead{Background} &
\colhead{$\rm\Gamma$} &
\colhead{$\rm Norm$} &
%\colhead{$N_{\rm H,uplim}$} &
%\colhead{C-stat/dof} &
\colhead{$\chi^2$/dof} &
\colhead{$P_{\rm null}$} &
%\colhead{$f_{\rm 2 keV}$}  &
% \colhead{$F_{\rm 0.5\textrm{--}2keV}$} &
\colhead{$L_{\rm X}$} \\%{$L_{\rm 2\textrm{--}10~keV}$} \\
% \colhead{   } &
%\colhead{ Date }&
%\colhead{(keV)}  &
%\colhead{Counts } &
%\colhead{Counts } &
% \colhead{    }  &
% \colhead{($10^{-5}$)}  &
%\colhead{($10^{13} \rm cm^{-2}$)} &
% \colhead{     }  &
% \colhead{     }  &
%\colhead{($\rm 10^{-32} erg~cm^{-2}~s^{-1}~Hz^{-1}$)} &
% \colhead{($\rm 10^{-15} erg~cm^{-2}~s^{-1}$) }  &
% s{-1}$)}\\
\colhead{(1)}         &
\colhead{(2)}         &
\colhead{(3)}         &
\colhead{(4)}         &
%\colhead{(5)}         &
\colhead{(5)}         &
\colhead{(6)}         
%\colhead{(8)}		  &
%\colhead{(9)}		  &
%\colhead{(10)}		  
%\colhead{(11)}		  &
% \colhead{(12)}		 
}
\startdata
%J0831 + 3213 & 2019--07--08 & $ 325$&$     1.89_{-0.16}^{+0.17}$& $3.52^{+0.84}_{-0.68}$ & $3.82$&$     234.6/231$& \\
SDSS J0831 + 3213 &$     1.99_{-0.23}^{+0.27}$& $3.42^{+1.30}_{-0.92}$ & $     16.4/19$&0.63&$9.55^{+1.48}_{-1.51}$\\
%$9.93_{-1.16}^{+0.51}$&$15.12_{-2.58}^{+3.71}$\\
SDSS J1423 + 0420 &$     1.86_{-0.14}^{+0.15}$& $2.99^{+0.63}_{-0.53}$ &$     40.8/49$&0.79&$9.87^{+1.09}_{-1.26}$
%$10.63_{-0.78}^{+0.50}$&$  16.73_{-3.01}^{+3.15}$&0.87&
\enddata
\tablecomments{
Col. (1): object name.
%Cols. (2): observation start date; 
%Cols. (3): X-ray energy band used for spectral fitting; 
% Cols. (3): total counts of the spectrum in the energy band shown in column (3); 
% Cols. (4): background counts of the spectrum in the energy band shown in column (3);
Col. (2): \hbox{power-law} photon index.
Col. (3): \hbox{power-law} normalization in units of $10^{-5}$ photons $\rm cm^{-2}~s^{-1}~keV^{-1}$.
% Cols. (7): upper limit of the column density of the possible X-ray absorber; 
Col. (4): $\chi^2$ value divided by the degrees of freedom.
%Cols. (9): observed-frame 2 keV flux in units of $\rm 10^{-32}erg\,cm^{-2}\,s^{-1}\,Hz^{-1}$;
Col. (5): null hypothesis probability of the model.
Col. (6): rest-frame 2–10 keV luminosity in units of $\rm 10^{43}~erg\,s^{-1}$.
}
\label{tab:spectra_fit}
\end{deluxetable*}

\begin{figure*}
\hspace{-0.2cm}
 \includegraphics[width=0.95\textwidth]{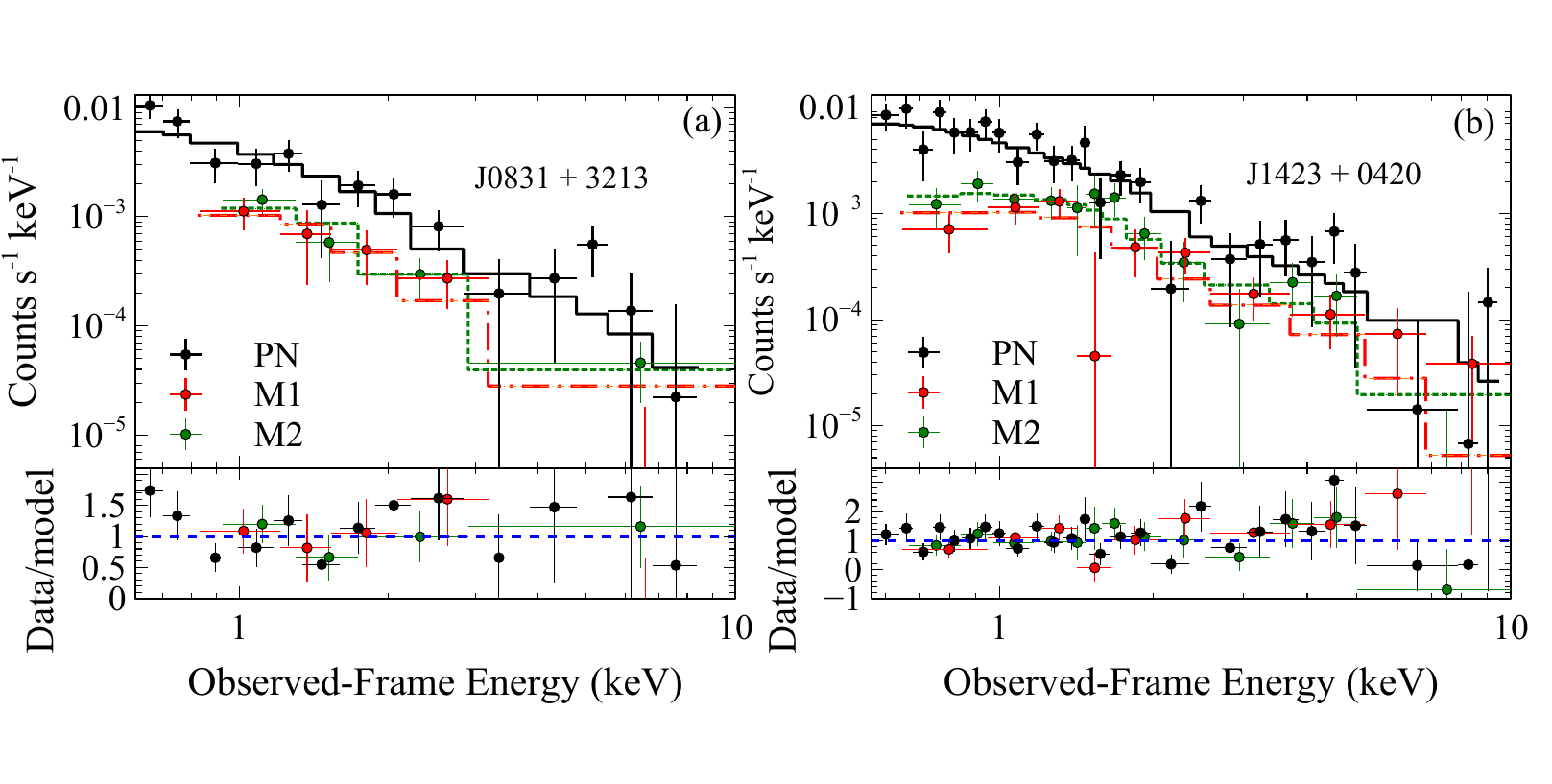}
 \centering
 \caption{The \hbox{XMM-Newton} spectra of (a) SDSS J0831 + 3213 and (b) SDSS J1423 + 0420 overlaid with the best-fit simple \hbox{power-law} models. The bottom panels show the ratios of the spectral data to the best-fit models. The EPIC pn (black), MOS1 (red), and MOS2 (green) spectra were jointly fitted.}
 %\caption{(a) The XMM spectra of SDSS J0831 + 3213 overlaid with the best-fit models. (b) The XMM spectra of SDSS J1423 + 0420 overlaid with the best-fit models. The spectra are grouped so that each bin has a detection at  $> 3\sigma$ level for display purposes. The joint fitting of the spectra taken by the three cameras (PN, MOS1, MOS2) is performed for the two sources. Both spectra sets can be fitted well by a simple \hbox{power-law} model.}
 \label{fig:xray_spectra}
\end{figure*}

\subsection{Optical spectra}
\label{subsec:simultaneous_opt_spectra}
An optical spectrum of SDSS J0831+3213 was taken by the Lijiang \hbox{2.4 m} telescope at the Yunnan Observatories of the Chinese Academy of Sciences on 2021 April 9 with an exposure time of 60 minutes. Optical spectra of SDSS J1423+0420 were obtained by the Hobby-Eberly Telescope (HET) at the McDonald Observatory on 2021 January 24 and by the Lijiang \hbox{2.4 m} telescope on 2021 February 9; the exposure times are both 50 minutes. Grism No.\ 3 was used in the Lijiang observations with a resolving power of $\approx2000$, and the data were reduced the same way as in \cite{2015ApJ...806...22D}. Flux calibration of the Lijiang spectra was carried out using the spectra of standard stars. For the HET observation of SDSS J1423+0420, we used the blue arm of the Low-Resolution Spectrograph 2 (LRS2-B) with a resolving power of $\approx1800$. The data were processed following the standard procedures using the HET pipeline tool panacea.\footnote{\url{https://github.com/grzeimann/Panacea}.} For absolute flux calibration, we normalized the HET spectrum to the Lijiang spectrum at rest-frame 2500 $\AA$ as the two observations are near simultaneous.

The Lijiang and HET (for SDSS J1423+0420) spectra of the two quasars are displayed in Figure \ref{fig:optical_spectra}. The spectra have been corrected for Galactic extinction using the \cite{2019ApJ...886..108F} Milky Way extinction model with $R_{V}$ = 3.1. The Galactic $E_{B-V}$ values of SDSS J0831+3213 and SDSS J1423+0420 are 0.0409 and 0.0237, respectively, obtained from \cite{2011ApJ...737..103S}. Since both quasars have \hbox{SDSS-I} spectra, which have good flux calibration in general \citep[e.g.,][]{2016ApJ...831..157M}, we show in Figure \ref{fig:optical_spectra} the corresponding SDSS spectra for comparison. The Lijiang and SDSS spectra of SDSS J0831+3213 agree well with each other. For SDSS J1423+0420, the near-simultaneous Lijiang and HET spectra appear redder than the SDSS spectrum, and they are $\approx$ 25\% brighter than the SDSS spectrum redward of $\approx 2000~\AA$. Such a long-term variability amplitude is not uncommon among quasars, but the spectral shape change is not consistent with the typical ``bluer when brighter'' quasar variability \citep[e.g.][]{1985ApJ...296..423C,2004ApJ...601..692V,2012ApJ...744..147S}. One explanation is that there is also a slight increase of the optical/UV extinction that makes the spectra redder. %\textcolor{red}{The broad absorption on the Mg II blue wing for SDSS J0831+3213 and the asymmetric Mg II emission line for SDSS J1423+0420 are caused by telluric absorption, since the reduction of Lijiang spectra do not correct the telluric absorption as done by the SDSS spectra reduction.} 
Nevertheless, the rest-frame UV spectra of the two quasars do not suggest any strong variability that could lead to the observed \hbox{X-ray} variability described in Section \ref{subsec:xray_variability} below. Both quasars show mild C IV blueshifts; for SDSS J0831+3213, the blueshift is $-1359\pm 962~\rm km~s^{-1}$, and for SDSS J1423+0420, it is $-652\pm 619~\rm km~s^{-1}$ \citep[][]{2022ApJS..263...42W}.

\section{X-ray and multiwavelength properties} \label{sec:xray-multiwavelength}
\subsection{X-ray spectral analysis}
\label{subsec:xray_spec_analysis}
%We used the $\chi^2$ statistic and C-statistic for the grouped and ungrouped spectra, respectively. 
The \hbox{XMM-Newton} spectra of SDSS J0831+3213 and SDSS J1423+0420 were fitted using XSPEC (v12.12.1, \citealt{arnaud1996astronomical}). We adopted a simple \hbox{power-law} model modified by Galactic absorption ({\sc zpowerlw*phabs}) to describe the 0.3--10 keV spectra. For each quasar, we jointly fitted the EPIC pn, MOS1, and MOS2 spectra, but we added normalization constants (best-fit values between 0.85 and 1.20) to the MOS spectra to allow for small cross-calibration uncertainties. The best-fit results are displayed in Table \ref{tab:spectra_fit} and Figure \ref{fig:xray_spectra}.

The simple \hbox{power-law} model describes the spectra well, with small reduced $\chi^{2}$ values and large null hypothesis probabilities (Table \ref{tab:spectra_fit}). The resulting \hbox{power-law} photon indices are $1.99^{+0.27}_{-0.23}$ and $1.86^{+0.15}_{-0.14}$, typical of type 1 quasars \citep[e.g.,][]{1997MNRAS.292..468R,2007ApJ...665.1004J,2011MNRAS.417..992S}. Adding an intrinsic absorption component ({\sc zphabs}) does not improve the fits, and we set upper limits on the intrinsic $N_{\rm H}$ of $4.1 \times 10^{21} \rm ~cm^{-2}$ and $6.6 \times 10^{21} \rm ~cm^{-2}$ for SDSS J0831+3213 and SDSS J1423+0420, respectively. From the best-fit models, we also computed rest-frame \hbox{2--10~keV} luminosities of the two quasars (Table \ref{tab:spectra_fit}), both approaching $10^{44} ~\rm erg~ s^{-1}$.

\subsection{X-ray variability}
\label{subsec:xray_variability}
Compared to the archival Chandra observations, the two quasars both turned out to be much brighter in the \hbox{XMM-Newton} observations. From the best-fit models, we calculated the rest-frame 2 keV flux densities ($f_{\rm 2 keV}$) and listed these in Table \ref{tab:Simultaneous_property}. We adopted the Chandra measurements of $f_{\rm 2 keV}$ from \cite{2020ApJ...900..141P}. The variability amplitudes in terms of the 2 keV flux densities are thus $4.0_{-1.4}^{+1.6}$ and $9.0_{-3.8}^{+7.4}$ for SDSS J0831 + 3213 and SDSS J1423 + 0420, respectively. Such large long-term variability amplitudes are rare among quasars \citep[e.g.,][]{1993ARA&A..31..717M,2012ApJ...746...54G,2017A&A...599A..82M,2020MNRAS.498.4033T}, which cannot be explained by typical quasar \hbox{X-ray} variability related to instability/fluctuations of the accretion disk and corona. Besides flux variability, the X-ray spectral shapes also appear different. In the recent XMM-Newton observations, both quasars had typical quasar spectral shapes ($\Gamma$ values of $1.99^{+0.27}_{-0.23}$ and $1.86^{+0.15}_{-0.14}$) with no signatures of X-ray absorption. For SDSS J0831+3213, the effective photon index from the Chandra observation is $\Gamma_{\rm eff}=1.0^{+0.6}_{-0.5}$, which is $\approx 1.6 \sigma$ smaller than the XMM-Newton $\Gamma$ value. For SDSS J1423+0420, the 90\% confidence-level upper limit on the effective photon index from the Chandra observation is $\Gamma_{\rm eff}<1.3$, still much smaller than the XMM-Newton $\Gamma$ value.

%As for the spectral shapes, the Chandra constraints with $\Gamma_{\rm eff}=1.0^{+0.6}_{-0.5}$ and $\Gamma_{\rm eff}<1.3$ suggest that the spectra were flatter than typical quasar spectra, indicative of \hbox{X-ray} absorption. In the recent \hbox{XMM-Newton} observations, both quasars had typical quasar spectral shapes ($\Gamma$ values of $1.99^{+0.27}_{-0.23}$ and $1.86^{+0.15}_{-0.14}$) with no signatures of \hbox{X-ray} absorption.
%We calculated the rest-frame 2 keV flux densities for the \hbox{XMM-Newton} observations of the two quasars based on the best-fit results (see Table \ref{tab:Simultaneous_property}). The calculated flux densities were compared with the Chandra 2 keV flux densities from \cite{2020ApJ...900..141P}, which are $1.03^{+0.35}_{-0.28} \times 10^{-32}$ \fluxd and $0.45^{+0.29}_{-0.20} \times 10^{-32}$ \fluxd for SDSS J0831 + 3213 and SDSS J1423 + 0420, respectively. The rest-frame 2 keV flux densities of the two quasars observed by \hbox{XMM-Newton} became much brighter than their Chandra observations. The 2 keV flux density variability factor is $4.0_{-1.4}^{+1.6}$ for SDSS J0831 + 3213 and it is $9.0_{-3.8}^{+7.4}$ for SDSS J1423 + 0420. The X-ray spectral shapes of the two quasars both change from the hard state observed by Chandra to the nearly flat spectra ($\Gamma \approx 2$) observed by \hbox{XMM-Newton}, indicating a transition from strong absorption to no absorption.

We also computed the $\alpha_{\rm OX}$ values for the two quasars to assess the \hbox{X-ray} emission strength relative to the optical/UV emission strength. For the \hbox{XMM-Newton} observations, we derived $f_{2500 \AA}$ and $L_{2500 \AA}$ values from the simultaneous Lijiang spectra, which are listed in Table~\ref{tab:Simultaneous_property}. For SDSS J0831 + 3213, the \hbox{XMM-Newton} OM photometric measurements also allow an estimate of $f_{2500 \AA}$ via extrapolation of an adopted \hbox{power-law} continuum with $a_{\rm \nu}=-0.46$ \citep[e.g.,][]{2001AJ....122..549V}. The resulting $f_{2500 \AA}$ value is consistent with that derived from the Lijiang spectrum. For the Chandra observations, since there were no simultaneous optical/UV measurements and both quasars do not show strong optical/UV variability (see Section \ref{subsec:SED_multiwavelength} below), we still adopted the same $f_{2500 \AA}$ and $L_{2500 \AA}$ values from the Lijiang spectra. We note that a 50\% difference in $f_{2500 \AA}$ only changes the resulting $\Delta \alpha_{\rm OX}$ value by a small amount of about 0.06.

\begin{figure}
\hspace{-0.2cm}
 \includegraphics[width=0.47\textwidth]{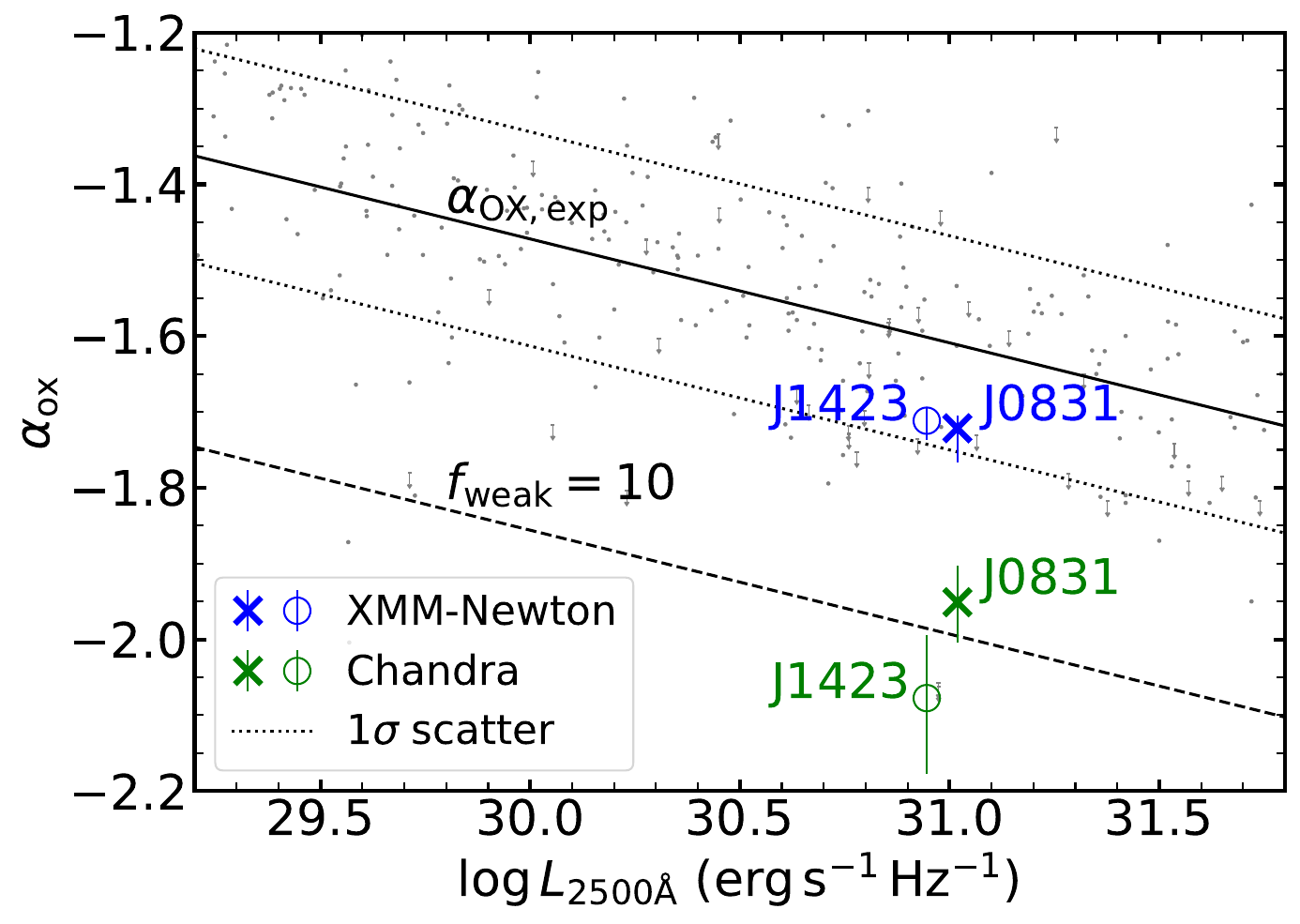}
 \centering
 \caption{ $\alpha_{\rm OX}$ vs.\ 2500 ${\rm \AA}$ monochromatic luminosity for the two quasars, displaying their significant \hbox{X-ray} weakness in the archival Chandra observations and nominal levels of \hbox{X-ray} emission in the \hbox{XMM-Newton} observations. The \cite{2006AJ....131.2826S} $\alpha_{\rm OX} \textrm{--} L_{\rm 2500 \AA}$ relation is shown as the solid line, with the dotted lines representing the $1 \sigma$ scatter of the relation at the luminosities of the two quasars. The dashed line indicates $f_{\rm weak}=10$ with respect to the $\alpha_{\rm OX} \textrm{--} L_{\rm 2500 \AA}$ relation. The small grey dots and downward arrows represent the $\alpha_{\rm OX}$ values and upper limits of the \cite{2006AJ....131.2826S} quasar sample, respectively. 
 }
 %with photon index $\Gamma > 1.26$.}
 \label{fig:alphaox}
\end{figure}

The $\alpha_{\rm OX}$ values, along with the corresponding $\Delta \alpha_{\rm OX}$ and $f_{\rm weak}$ values derived using the \cite{2006AJ....131.2826S} $\alpha_{\rm OX} \textrm{--} L_{\rm 2500 \AA}$ relation, are listed in Tables \ref{tab:Simultaneous_property}. We also show the two quasars in the $\alpha_{\rm OX}$ versus $L_{\rm 2500 \AA}$ plane in Figure \ref{fig:alphaox}. Both quasars were significantly \hbox{X-ray} weak in the Chandra observations. Given the $1\sigma$ scatter ($\Delta \alpha_{\rm OX}=0.14$; Table 5 of S06) of the \cite{2006AJ....131.2826S} $\alpha_{\rm OX} \textrm{--} L_{\rm 2500 \AA}$ relation, these $f_{\rm weak}$ values correspond to 2.4$\sigma$ and 3.4$\sigma$ deviations from the $\alpha_{\rm OX} \textrm{--} L_{\rm 2500 \AA}$ relation, respectively. However, both quasars recovered to nominal levels of \hbox{X-ray} emission in the \hbox{XMM-Newton} observations with $\Delta \alpha_{\rm OX}$ values of $\approx -0.11$, within the $1 \sigma$ scatter of the \cite{2006AJ....131.2826S} $\alpha_{\rm OX} \textrm{--} L_{\rm 2500 \AA}$ relation. Considering also the flatter spectral shapes in the Chandra observations, the \hbox{X-ray} weak states were likely not intrinsic, but they were simply caused by \hbox{X-ray} obscuration. In the recent \hbox{XMM-Newton} observations, there is no \hbox{X-ray} obscuration, and thus both the spectral shapes and flux levels return to those of typical quasars.

%With $f_{\rm weak}$ = $7.7^{+2.9}_{-1.9}$ and $17.4^{+14.2}_{-6.8}$, they were the 2.4$\sigma$ and 3.4$\sigma$ outliers of the \cite{2006AJ....131.2826S} $\alpha_{\rm OX} \textrm{--} L_{\rm 2500 \AA}$ relation, respectively, when observed by Chandra. 

%The estimated 2500 $\rm \AA$ and 2 keV flux densities, as well as the derived $\alpha_{\rm OX}$ parameters, are given in Table \ref{tab:Simultaneous_property}. The corresponding $\Delta \alpha_{\rm OX}$ and $f_{\rm weak}$ parameters were also computed using the \cite{2006AJ....131.2826S} $\alpha_{\rm OX} \textrm{--} L_{\rm 2500 \AA}$ relation. The $\alpha_{\rm OX}$ versus $L_{\rm 2500 \AA}$ distribution of the two quasars are shown in Figure \ref{fig:alphaox}. For the Chandra observation, SDSS J0831 + 3213 and SDSS J1423 + 0420 are significantly \hbox{X-ray} weak with $f_{\rm weak}$ values of $9.52^{+3.62}_{-2.41}$ and $22.41^{+18.32}_{-8.82}$, respectively. However, they change from the \hbox{X-ray} weak state to the nearly \hbox{X-ray} nominal-strength state when observed by \hbox{XMM-Newton}, both with $\Delta \alpha_{\rm OX}$ values of $-0.11$, which is within the $1\sigma$ scatter of the \cite{2006AJ....131.2826S} $\alpha_{\rm OX} \textrm{--} L_{\rm 2500 \AA}$ relation.

\begin{figure*}
\hspace{-0.2cm}
 \includegraphics[width=0.95\textwidth]{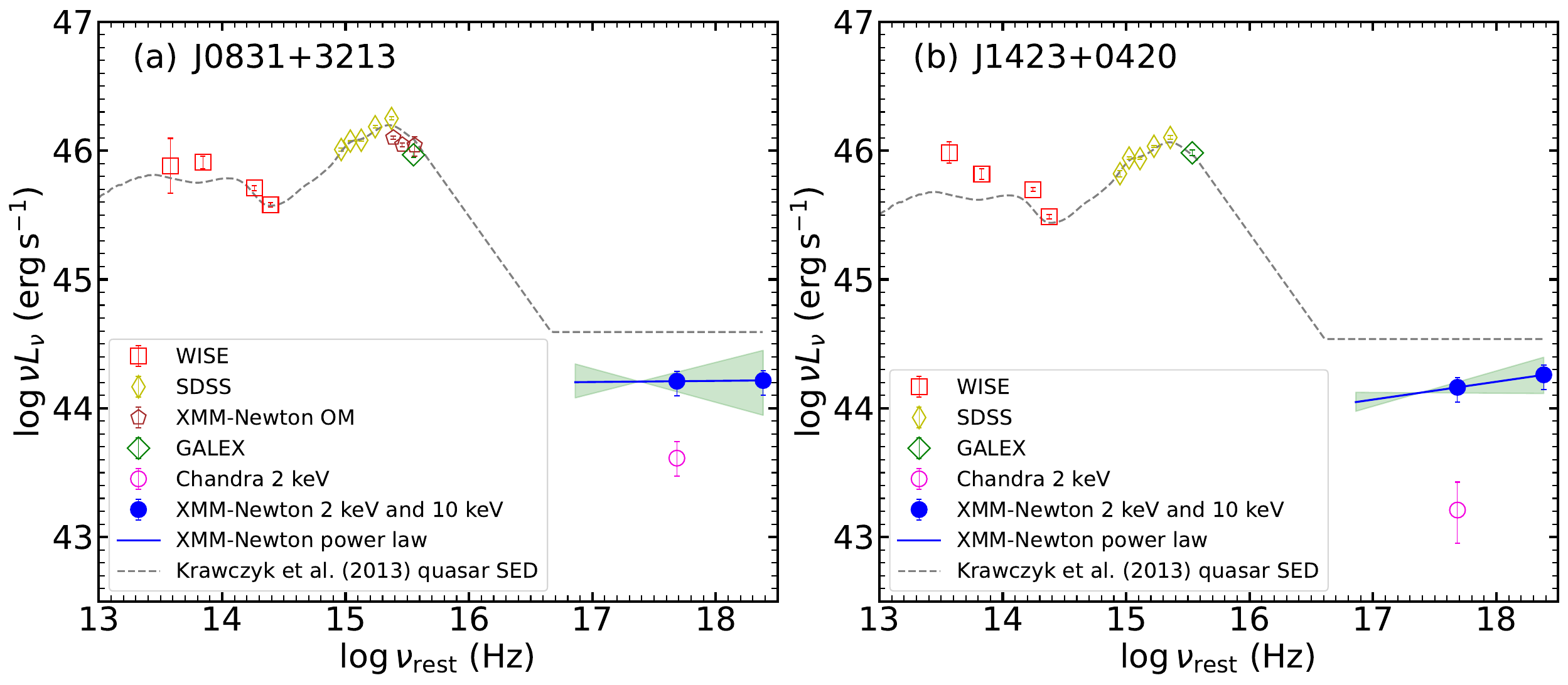}
 \centering
 \caption{IR-to-X-ray SEDs of (a) SDSS J0831 + 3213 and (b) SDSS J1423 + 0420. For SDSS J0831 + 3213, we also show the \hbox{XMM-Newton} OM measurements. The \hbox{XMM-Newton} 2 keV and 10 keV luminosities were derived from the best-fit models (Section \ref{subsec:xray_spec_analysis}). The \hbox{X-ray} spectral slopes and their uncertainties are indicated by the blue lines and the green-shaded areas. The Chandra 2 keV luminosities were adopted from \cite{2020ApJ...900..141P}. The grey dashed curve in each panel shows the mean quasar SED from \cite{2013ApJS..206....4K} normalized to the 2500 $\AA$ luminosity; the \hbox{X-ray} component is a $\Gamma$ = 2 \hbox{power-law} continuum with the 2 keV luminosity determined from the \cite{2006AJ....131.2826S} $\alpha_{\rm OX} \textrm{--} L_{\rm 2500 \AA}$ relation ($f_{\rm weak} = 0$).}
 \label{fig:SED}
\end{figure*}

\subsection{Spectral energy distributions and optical--IR light curves} \label{subsec:SED_multiwavelength}
We constructed IR-to-X-ray SEDs for the two quasars, shown in Figure \ref{fig:SED}. The IR--UV photometric data were collected from the Wide-field Infrared Survey Explorer \citep[WISE;][]{2010AJ....140.1868W,WISEAllSkySourceCatalog}, SDSS \citep[][]{2000AJ....120.1579Y}, and Galaxy Evolution Explorer \citep[GALEX;][]{2005ApJ...619L...1M} catalogs. The GALEX data presented in this paper were obtained from the Mikulski Archive for Space Telescopes (MAST) at the Space Telescope Science Institute. The specific observations analyzed can be accessed via \dataset[doi:10.17909/T9H59D]{https://doi.org/10.17909/T9H59D}. For these two quasars, only near-UV (NUV) measurements are available in the GALEX catalog; the non-detections in the far-UV (FUV) band are probably caused by the Lyman break in these $z\approx 1.7$ quasars. All the SED data have been corrected for the Galactic extinction. For SDSS J0831 + 3213, the \hbox{XMM-Newton} OM measurements were added to the SED plot. For both quasars, we added the \hbox{XMM-Newton} 2 keV and 10 keV luminosities determined from the best-fit results in Section \ref{subsec:xray_spec_analysis}. We also show the \hbox{X-ray} spectral slopes and their uncertainties. For the Chandra observations, we added the 2 keV luminosities from \cite{2020ApJ...900..141P}. For comparison, we include the mean SED of high-luminosity quasars in \cite{2013ApJS..206....4K}, scaled to the 2500 $\AA$ luminosities of our objects. An \hbox{X-ray} component was added to the mean SED to indicate the \cite{2006AJ....131.2826S} $\alpha_{\rm OX} \textrm{--} L_{\rm 2500 \AA}$ relation. We note that the IR--UV SED data are not simultaneous, and they may be affected by mild variability (see light curves below). However, the IR--UV SEDs of both quasars are still broadly consistent with those of typical quasars. SDSS J1423 + 0420 shows somewhat stronger mid-IR emission, which is not unusual considering the complex quasar IR emission mechanisms \citep[e.g.,][]{2017ApJ...835..257L,2018ApJ...862..118Z}. 

\begin{figure*}
\hspace{-0.2cm}

 \centering
 \includegraphics[trim=0 0 0 0,clip, width=0.82\linewidth]{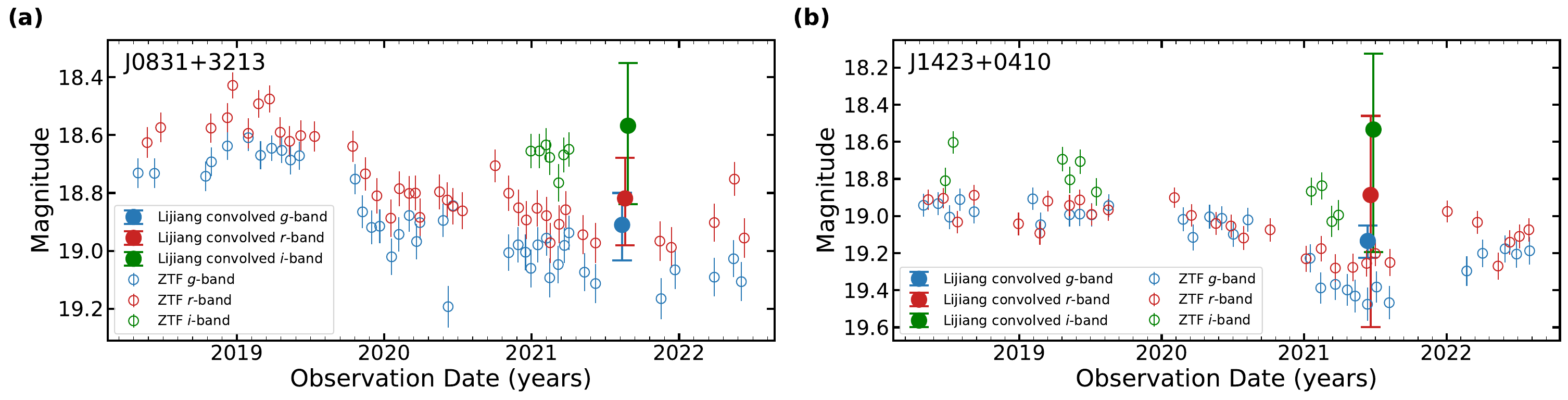}
 \includegraphics[trim=0 0 0 0,clip, width=0.82\linewidth]{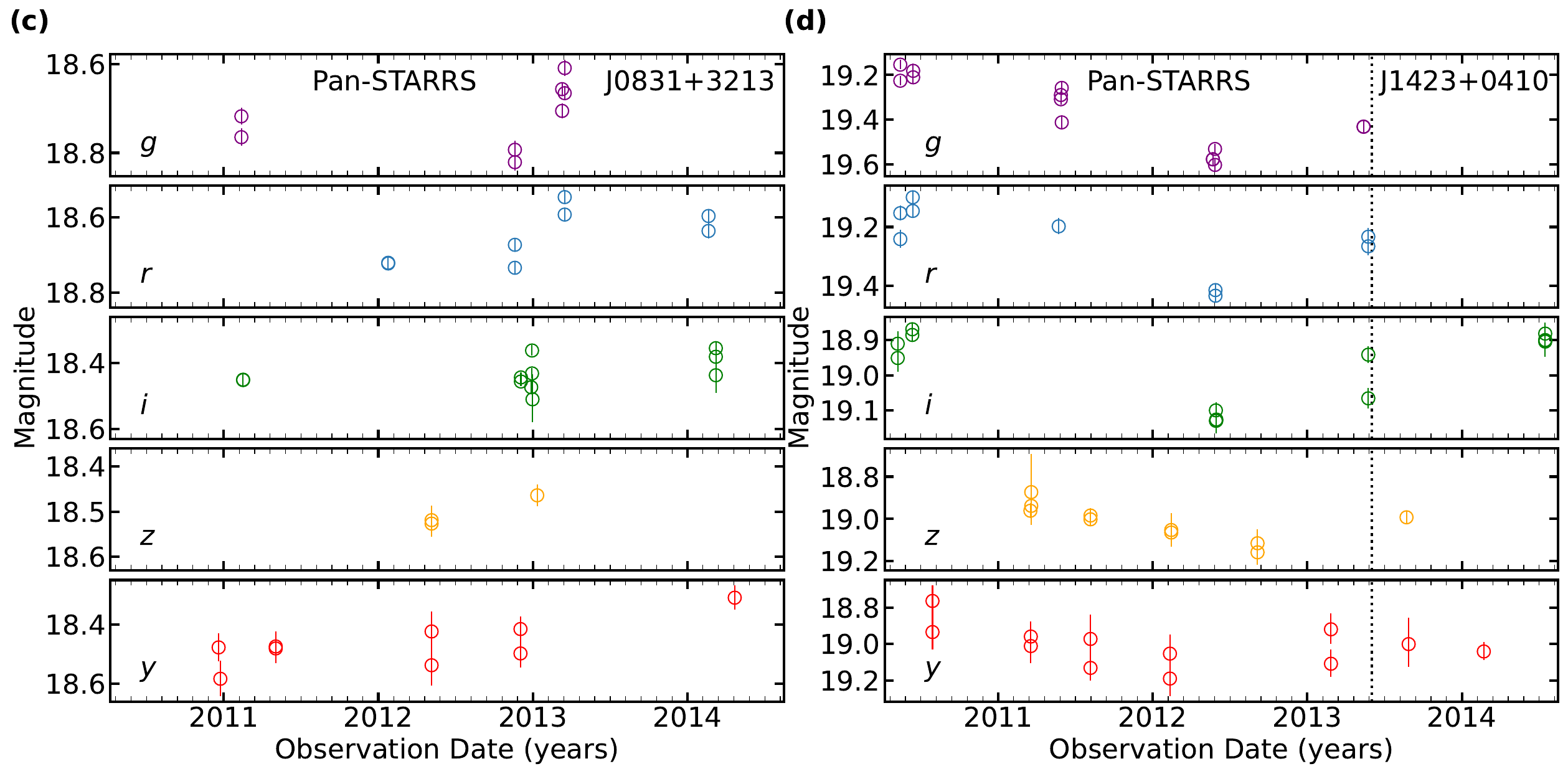}
 \includegraphics[trim=0 0 10 0,clip, width=0.82\linewidth]{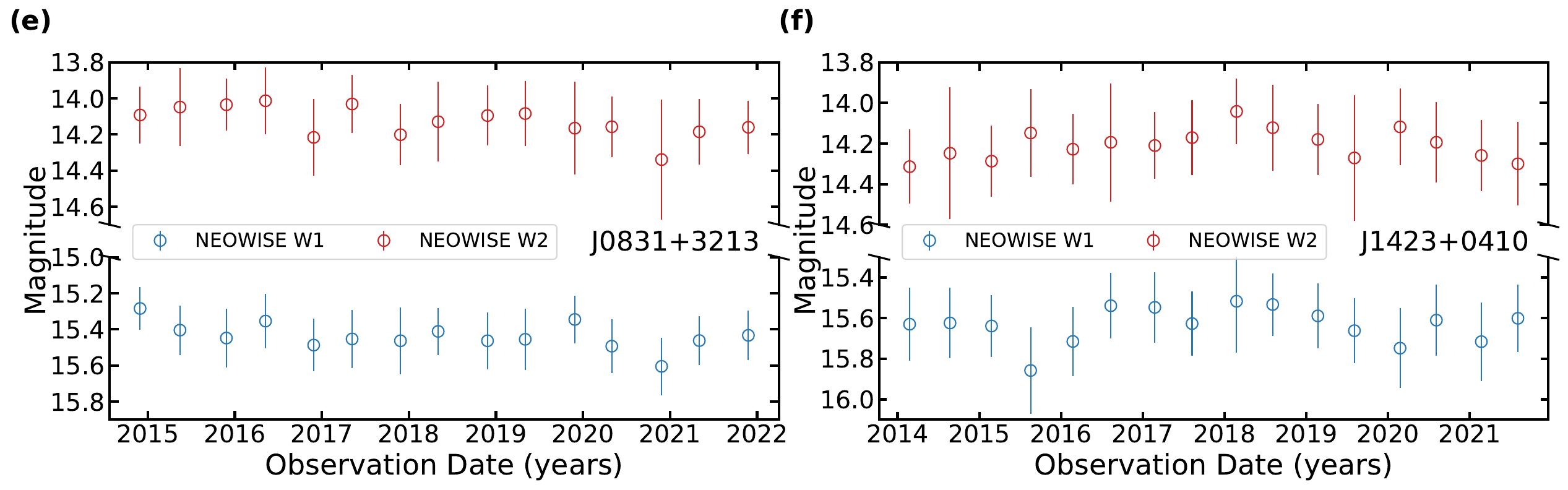}
 \caption{(a), (b) ZTF $g$-, $r$-, and $i$-band light curves of SDSS J0831 + 3213 and SDSS J1423 + 0410. (c), (d) Pan-STARRS $g$-, $r$-, $i$-, $z$-, and $y$-band light curves. (e), (f) NEOWISE W1- and W2-band light curves. All the data points have been corrected for the Galactic extinction. In the ZTF light curves, we added corresponding measurements from the Lijiang spectra, which are broadly consistent with the ZTF data. The vertical dotted lines in the Pan-STARRS light curves of SDSS J1423 + 0410 represent the date of its Chandra observation. For the NEOWISE light curves, we grouped any intra-day measurements. These light curves indicate that the two quasars do not show any substantial long-term variability in the optical/UV and IR bands. }

% (a), (b) The ZTF g, r, and i band light curves of SDSS J0831 + 3213 and SDSS J1423 + 0410. (c), (d) The NEOWISE W1, W2 band light curves of SDSS J0831 + 3213 and SDSS J1423 + 0410. All the data points have been corrected for Galactic extinction. For the NEOWISE light curves, we grouped any intra-day measurements}
 \label{fig:lightcurve}
\end{figure*}

To investigate the IR--UV photometric variability of the two quasars, we collected their multi-epoch data from the Zwicky Transient Facility \citep[ZTF;][]{2019PASP..131a8003M},  Panoramic Survey Telescope and Rapid Response System \citep[Pan-STARRS;][]{2020ApJS..251....7F}, and Near-Earth Object Wide-field Infrared Survey Explorer Reactivation Mission \citep[NEOWISE;][]{2011ApJ...731...53M,neowise} catalogs. The Pan-STARRS data presented in this paper were obtained from the MAST at the Space Telescope Science Institute. The specific observations analyzed can be accessed via \dataset[doi:10.17909/s0zg-jx37]{https://doi.org/10.17909/s0zg-jx37}. The ZTF, Pan-STARRS, and NEOWISE light curves are presented in Figure \ref{fig:lightcurve}. The ZTF light curves overlap the dates of the Lijiang observations, and thus we derived corresponding photometric measurements from the Lijiang spectra and added these to the ZTF light curves. For SDSS J0831 + 3213, the Lijiang measurements are slightly brighter than the adjacent ZTF data, but they are separated by a few months. For SDSS J1423 + 0410, the Lijiang measurements are consistent with the ZTF data within the uncertainties. The flux calibration of the Lijiang spectra thus appears reasonable.

These optical/UV and IR light curves indicate that the two quasars do not show any substantial long-term variability in these bands. The maximum variability amplitude reaches $\approx 0.5$ mag (a factor of $\approx$ 1.6) in these light curves, still much smaller than the \hbox{X-ray} variability factors (Section \ref{subsec:xray_variability}). Therefore, the \hbox{X-ray} weakness and \hbox{X-ray} variability of these two quasars are not likely connected to any changing-look behavior where the optical/UV and \hbox{X-ray} continua vary coordinately. 

%We also presented the ZTF g, r, and i bands magnitudes and uncertainties of the two quasars convolved from the Lijiang spectra. The Lijiang observation of SDSS J0831 + 3213 does not overlap with the ZTF light curve, and it appears that there may be a mild variability between the Lijiang observation and the ZTF photometry measurement. For SDSS J1423 + 0410, the convolved Lijiang magnitudes are consistent with the ZTF photometry measurement within the uncertainties. The mild variability in rest-frame IR and optical bands indicate that the accretion powers of the two quasars did not change significantly. Thus the X-ray variability and state transition of the two quasars should not be caused by the potential strong UV variability.

\section{Discussion and Summary}
\label{sec:summary_discussion}
The \hbox{XMM-Newton} observations of the two quasars reveal that they have recovered to nominal levels of \hbox{X-ray} emission with typical \hbox{X-ray} spectral shapes (Sections \ref{subsec:xray_spec_analysis} and \ref{subsec:xray_variability}). The \hbox{X-ray} weak states in the Chandra observations that motivated this study are thus likely caused by \hbox{X-ray} obscuration considering the flat spectral shapes ($\Gamma_{\rm eff}$ = $1.0^{+0.6}_{-0.5}$ and $\Gamma_{\rm eff} < 1.3$). Given the variability factors ($\approx$ 4.0 and 9.0), we estimate that the column densities of the absorbers are $\approx 2.5 \times \rm 10^{22}~cm^{-2}$ and $\approx 4.8 \times \rm 10^{22}~cm^{-2}$ by adding an intrinsic absorption component to the best-fit models in Section \ref{subsec:xray_spec_analysis}. The \hbox{X-ray} variability between the Chandra and \hbox{XMM-Newton} observations is then explained with changes of \hbox{X-ray} obscuration.

Of the 426 sample B $+$ C quasars in \citet{2020ApJ...900..141P}, 14 ($ \approx$ 3.3\%) quasars were identified as the unclassified type of \hbox{X-ray} weak quasars which have typical quasar UV continua and emission-line spectra. Out of these 14 objects, we selected two targets and obtained just one additional \hbox{XMM-Newton} observation for each. Yet they both turned out to be typical quasars considering their \hbox{X-ray} properties, UV spectra, and SEDs. These results suggest that the \hbox{X-ray} weak states caught in the archival Chandra observations are rare, and these quasars should show nominal levels of \hbox{X-ray} emission most of the time. This is also consistent with the small fraction of such objects found in the \citet{2020ApJ...900..141P} survey study. In this case, if we reobserve the other 12 unclassified type of \hbox{X-ray} weak quasars in \citet{2020ApJ...900..141P}, we should find nominal levels of \hbox{X-ray} emission in most of them. It is even probable that typical quasars might also become \hbox{X-ray} obscured occasionally (e.g., $\approx 3.3 \%$ of the time) if long-term \hbox{X-ray} monitoring observations are available.

Since the long-term optical/UV and IR light curves do not show substantial variability (Section \ref{subsec:SED_multiwavelength}), and the multi-epoch UV spectra do not vary much (Section \ref{subsec:simultaneous_opt_spectra}), the \hbox{X-ray} absorber likely does not affect the optical/UV continuum emission. A good candidate for such an absorber is the small-scale clumpy accreting-disk wind that is dust free. We cannot exclude the possibility that unusual UV spectral features accompanied the Chandra \hbox{X-ray} weak states. For example, they might have exhibited BAL features during the Chandra observations (Section \ref{sec:intro}), and they showed \hbox{non-BAL-to-BAL-to-non-BAL} transitions among the SDSS (non-BAL), Chandra (assumed BAL), and HET/Lijiang (non-BAL) observations, which are separated by 1.9--4.1 years in the rest frame. However, the probability of having such multiple BAL transitions appears insufficient to explain the 3.3\% fraction of the unclassified type of X-ray weak quasars in \citet{2020ApJ...900..141P}. For example, \citet{2018ApJ...862...22R} found a non-BAL-to-BAL quasar emergence rate of $0.59\% \pm 0.12\%$ based on a parent sample of $\approx 15\,000$ SDSS quasars; among these BAL transition cases, only about 1/6 of the objects exhibited both emergence and disappearance. Nevertheless, absorption from the clumpy disk wind is still a probable cause of the \hbox{X-ray} weakness in BAL quasars \cite[e.g.,][]{2016MNRAS.458..293M}.
%For example, they might have exhibited BAL features during the Chandra observations (Section \ref{sec:intro}) and belong to the rare population of quasars showing emerging and/or disappearing BALs \cite[e.g.,][]{2012ApJ...757..114F,2017MNRAS.469.3163M,2018A&A...616A.114D,2018ApJ...862...22R,2019MNRAS.482.1121S,2021ApJS..255...12Y}. The SDSS (non-BAL), Chandra (assumed BAL), and HET/Lijiang (non-BAL) observations are separated by 1.9--4.1 years in the rest-frame. 

%However, the emerging/disappearing rate of BALs in BALs quasars population is $\sim 5\%$ \cite[e.g.,][]{2018A&A...616A.114D,2018ApJ...862...22R}, so the fraction of such sources can not account for the occurrence rate of the unclassified type of X-ray weak quasars considering that BAL quasar fraction in the optical survey is $\sim 15\%$ \citep[e.g.,][]{2009ApJ...692..758G}. And even in this case, absorption from the clumpy disk wind is still a probable cause of the \hbox{X-ray} weakness \cite[e.g.,][]{2016MNRAS.458..293M}. 

Obscuration from the disk wind (and from a possible geometrically thick accretion disk) has been suggested to explain the \hbox{X-ray} weakness observed in some WLQs and super-Eddington accreting quasars \cite[e.g.,][]{2012MNRAS.425.1718M,2015ApJ...805..122L,2020ApJ...889L..37N,2022MNRAS.511.5251N,2021ApJ...910..103L,2022ApJ...930...53L,2023ApJ...950...18H}. Powerful disk winds launched via radiation pressure are expected in these quasars \cite[e.g.,][]{2014ApJ...796..106J,2019ApJ...880...67J,2014MNRAS.439..503S}, and they could potentially provide significant \hbox{X-ray} obscuration. SDSS J0831 + 3213 and SDSS J1423 + 0410 do not show weak UV emission lines like WLQs, and their estimated Eddington ratios are not high (Table \ref{tab:objectproperty}). Therefore, they might simply be typical type 1 quasars that occasionally develop powerful winds and enter the rare X-ray weak/absorbed states. Besides the wind density and covering factor, the probability of observing such extreme X-ray variability among typical quasars should also have an orientation dependence, as a large inclination angle is preferred for observing X-ray obscuration from the equatorial disk wind. Identification of more such objects is required to clarify their nature.

%Therefore, they might simply be typical quasars that occasionally develop a high-density wind and enter the rare \hbox{X-ray} weak/absorbed states. \textbf{As such disk wind may not have large opening angle and it can only cause \hbox{X-ray} obscuration when observed with a relatively equatorial viewing angle, the extreme \hbox{X-ray} variability of this type of Type 1 quasars might be orientation dependent.} Identification of more such objects is required to clarify their nature.

%\textbf{The extreme X-ray variability of this type of Type 1 quasars might be orientation dependent, as the disk wind with limited opening angle can only cause X-ray obscuration when observed with equatorial viewing angle.} 

In summary, we present the \hbox{XMM-Newton} observations of two examples of an unclassified type of \hbox{X-ray} weak quasars from the \citet{2020ApJ...900..141P} Chandra survey, SDSS J0831 + 3213 and SDSS J1423 + 0410. Their UV continua and emission-line spectra are similar to typical quasars and lack BALs, and they have typical quasar IR--UV SEDs. In the \hbox{XMM-Newton} observations, both quasars show nominal levels of \hbox{X-ray} emission with typical quasar \hbox{X-ray} spectral shapes, displaying strong \hbox{X-ray} variability compared to the archival Chandra data. Simultaneous optical (rest-frame UV) spectra indicate no strong variability compared to the SDSS spectra. Long-term optical/UV and IR light curves do not show any substantial variability either. We consider that the \hbox{X-ray} weakness observed in the Chandra observations is due to \hbox{X-ray} obscuration from a small-scale dust-free absorber, likely related to accretion-disk winds. Such \hbox{X-ray} weak/absorbed states are probably rare in typical quasars like SDSS J0831 + 3213 and SDSS J1423 + 0410, and thus they both recovered to \hbox{X-ray} \hbox{nominal-strength} states in the \hbox{XMM-Newton} observations. Future observations of similar objects (e.g., the other 12 unclassified type of \hbox{X-ray} weak quasars in \citealt{2020ApJ...900..141P}) should be able to provide constraints on the duty cycles of the \hbox{X-ray} weak states in these quasars and thus clarify their nature. We note that even in the high states, the expected 0.5--2 keV fluxes of these \citet{2020ApJ...900..141P} quasars derived from the $\alpha_{\rm OX} \textrm{--} L_{\rm 2500 \AA}$ relation are still $\approx$ 5--10 times lower than the sensitivity limit of the eROSITA survey \citep[e.g.,][]{2012arXiv1209.3114M,2021A&A...647A...1P}, and thus targeted observations by XMM-Newton/Chandra are needed.

%The high state flux of these quasars is about 5--10 times lower than the typical eROSITA flux limit in the 0.5--2.0 keV energy band \citep{2012arXiv1209.3114M,2021A&A...647A...1P}, so they can not be detected in the future eROSITA surveys and require target observations by XMM-Newton/Chandra.

\begin{deluxetable*}{ccccccccccc}
%\begin{table}[t]{cccccccccccc}
\tabletypesize{\scriptsize}
%\tabletypesize{\small}
\tablewidth{0.2pt}
\tablecaption{\hbox{X-ray} and Optical/UV Properties
}
\tablehead{
\colhead{Object Name}&
\colhead{Observatory}&
\colhead{Date} &
\colhead{$f_{\rm 2keV}$} &
\colhead{$f_{\rm 2500 \AA}$} &
\colhead{$L_{\rm 2500 \AA}$} &
% \colhead{$f_{\rm UVM2}$} &
% \colhead{$f_{\rm UVW1}$} &
% \colhead{$f_{\rm U}$} &
\colhead{$\alpha_{\rm OX}$} &
\colhead{$\Delta \alpha_{\rm OX}$} &
\colhead{$f_{\rm weak}$}  \\
\colhead{(1)}         &
\colhead{(2)}         &
\colhead{(3)}         &
\colhead{(4)}         &
\colhead{(5)}         &
\colhead{(6)}         &
\colhead{(7)}         &
\colhead{(8)}		  &
\colhead{(9)}		  
% \colhead{(10)}		  &
% \colhead{(11)}		 
}
\startdata
% SDSS J0831 + 3213 & Chandra & 2007--12--22 &$ 1.03^{+0.35}_{-0.28}$& $1.25$&$1.04$&$     -1.95^{+0.05}_{-0.05}$&$-0.38$&$9.52^{+3.62}_{-2.41}$\\
% $ $& \hbox{XMM-Newton} & 2019--07--08 &$ 4.08^{+0.42}_{-0.96}$& $1.25$&$1.04$&$     -1.71^{+0.02}_{-0.02}$&$-0.13$&$2.19^{+0.23}_{-0.20}$\\
% SDSS J1423 + 0420  & Chandra & 2012--12--15&$0.45^{+0.29}_{-0.20}$&$1.16 $&$0.88$&$-2.08^{+0.08}_{-0.10}$&$-0.52$&$22.41^{+18.32}_{-8.82}$\\
% $ $& \hbox{XMM-Newton}& 2021--04--07&$4.04^{+0.41}_{-0.57}$&$1.16 $&$0.88$&$-1.70^{+0.01}_{-0.01}$&$-0.14$&$2.33^{+0.22}_{-0.15}$ 
SDSS J0831 + 3213 & Chandra & 2007--12--22 &$ 1.03^{+0.35}_{-0.28}$& $1.25$&$1.04$&$     -1.95^{+0.05}_{-0.05}$&$-0.34$&$7.7^{+2.9}_{-1.9}$\\
$ $& \hbox{XMM-Newton} & 2021--04--07 &$ 4.08^{+0.42}_{-0.96}$& $1.25$&$1.04$&$     -1.72^{+0.02}_{-0.04}$&$-0.11$&$1.9^{+0.6}_{-0.2}$\\
SDSS J1423 + 0420  & Chandra & 2012--12--15&$0.45^{+0.29}_{-0.20}$&$1.16 $&$0.88$&$-2.08^{+0.08}_{-0.10}$&$-0.48$&$17.4^{+14.2}_{-6.8}$\\
$ $& \hbox{XMM-Newton}& 2021--01--25&$4.04^{+0.41}_{-0.57}$&$1.16 $&$0.88$&$-1.71^{+0.02}_{-0.03–}$&$-0.11$&$1.9^{+0.3}_{-0.2}$ 
\enddata
\tablecomments{
Col. (1): object name.
Col. (2): observatory. 
Col. (3): observation date.
Col. (4): rest-frame 2 keV flux density in units of $10^{-32}$ $\rm erg~cm^{-2}~s^{-1}~Hz^{-1}$; the errors of the Chandra measurement were propagated from the count errors. 
Col. (5): rest-frame 2500 $\rm \AA$ flux density in units of $10^{-27}$ $\rm erg~cm^{-2}~s^{-1}~Hz^{-1}$, derived from the Lijiang spectrum.
Col. (6): 2500 $\rm \AA$ monochromatic luminosity in units of $10^{31}$ $\rm erg~s^{-1}~Hz^{-1}$.
% Cols. (6)-(8): the Galactic extinction corrected UV/optical flux densities in the OM bands, in units of $10^{-27}$ $\rm erg~cm^{-2}~s^{-1}~keV^{-1}$; 
Col. (7): X-ray-to-optical \hbox{power-law} slope parameter with the uncertainties propagated from the $f_{\rm 2keV}$ uncertainties.
Col. (8): difference between the observed $\alpha_{\rm OX}$ value and the expected $\alpha_{\rm OX}$ value derived from the $\alpha_{\rm OX}$ - $L_{\rm 2500 \AA}$ relation of \cite{2006AJ....131.2826S}. 
Col. (9): X-ray weakness factor: $f_{\rm weak} = 10^{-\Delta \alpha_{\rm OX}/0.384}$.
}
\label{tab:Simultaneous_property}
\end{deluxetable*}

\begin{acknowledgments}

We thank Donald P. Schneider and Sergey Rostopchin for help with the HET observation. We acknowledge the support of the staff of the Lijiang 2.4 m telescope. Z.Z. and B.L. acknowledge financial support from the National Natural Science Foundation of China grant 11991053, China Manned Space Project grants \hbox{NO. CMS-CSST-2021-A05} and \hbox{NO. CMS-CSST-2021-A06}. W.N.B. acknowledges support from the Eberly Chair Endowment at Penn State. J.M.W. acknowledges financial support from the National Natural Science Foundation of China grants NSFC-11991050, -11991054, and -11833008.

The HET is a joint project of the University of Texas at Austin, \hbox{Ludwig-Maximillians-Universität} München, \hbox{Georg-August-Universität} Göttingen, and the Pennsylvania State University. Funding for the Lijiang 2.4 m telescope has been provided by the Chinese Academy of Sciences and the People’s Government of Yunnan Province.

%The HET is named in honor of its principal benefactors, William P.Hobby and Robert E. Eberly. The LRS2 was developed and funded by the University of Texas at Austin McDonald Observatory and Department of Astronomy, and by the Pennsylvania State University.

%W.N.B. thanks the Eberly environment at Pennsylvania State University.

%We acknowledge the support of the staff of the Lijiang 2.4 m telescope (LJT). 
\end{acknowledgments}

%\bibliography{ms}{}
%\bibliographystyle{aasjournal}

\end{document}